\documentclass[fleqn,twoside]{article}
\usepackage{espcrc2}

\usepackage{graphicx}
\usepackage[figuresright]{rotating}

\newcommand{\AmS}{{\protect\the\textfont2
  A\kern-.1667em\lower.5ex\hbox{M}\kern-.125emS}}

\title{Nonequilibrium Aspects of Quantum Field Theory}
\author{Travis R. Miller
	\address[stupidaddressmark] {Dept. of Physics, 
        Washington University, \\ 
        St. Louis, MO 63130 USA}%
        \thanks{We gratefully acknowledge the support of the U.S. Dept. of
		Energy under DOE DE-FG02-91ER40628},
        Michael C. Ogilvie \addressmark[stupidaddressmark]}

\begin{document}

\begin{abstract}
We have developed a method for extracting equilibrium observables from non-equilibrium
simulations by rapidly changing the temperature and recording the subsequent
evolution of the Polyakov loop. Both nucleation and spinodal decomposition are
observed to occur.  In the latter case the Polyakov loop correlation function shows 
exponential growth for wavenumbers less than or equal to the critical wavenumber 
\( k_{c} \).  We have constructed the bare as wee as the effective potential for 
the Polyakov loop, from which \( k_{c} \) and $m_D/k_c$ can be extracted as a 
function of temperature. The shift from spinodal decomposition to nucleation as the 
dominant equilibration mechanism occurs at the spinodal temperature that separates 
these two regimes.
\vspace{1pc}
\end{abstract}

\maketitle

\section{Introduction}

Spinodal decomposition is characterized by the appearance of 
explosive growth in the low momentum modes of the system, while the high momentum 
modes relax to their equilibrium distributions. The critical wavenumber 
\( k_{c} \) which separates low and high momentum behaviors is determined by 
the thermodynamic potential which governs the long wavelength behavior of the system.
We have previously shown that the confined phase of pure $SU(2)$ gauge theory decays via 
spinodal decomposition after a rapid change from below the 
deconfinement temperature to temperatures above\cite{us}.

The first order character of the deconfinement transition in pure $SU(3)$ gauge
theory implies that the confined phase is metastable for a range of temperatures
above the deconfinement temperature $T_d.$  At higher temperatures $T>T_s>T_d$ the
confined phase is unstable and decays via spinodal decomposition.  The spinodal
temperature $T_s$ separates the metastable and spinodal temperature ranges.

\section{Dynamics of the Polyakov Loop}

As a consequence of asymptotic freedom, the perturbative form of the effective
potential for the Polyakove loop at high temperatures is valid. Writing the
Polyakov loop in the form \( P_{F}=1+2cos(2\pi /3-\psi ) \), the perturbative
form of the effective action is at 1 loop \cite{gross}\cite{weiss}\cite{bhat}
\begin{eqnarray}
S_{eff}=\frac{4T^{2}}{g^{2}}\int d^{3}x ( \frac{1}{2}(\nabla \psi )^{2}-
\frac{1}{2}(\frac{g^{2}T^{2}}{3})\psi^{2} \nonumber \\
-\frac{g^{2}T^{2}}{6\pi }\psi^{3}+\frac{3g^{2}T^{2}}{8\pi^{2}}\psi^{4} ),
\end{eqnarray}
where the constant black body term has been discarded. 

We assume that the long distance behavior of the Polyakov loop 2-point function 
can be calculated using the 1-loop perturbative effective action \( S_{eff} \) 
and the Langevin equation,
\begin{equation}
\frac{\partial P}{\partial t}=-\Gamma \frac{\delta S_{eff}}{\delta P^{*}}+\eta 
\end{equation}
where \( \eta  \), the Gaussian noise term, is such that the equipartition
theorem holds\cite{us}.
A linearized solution to the Langevin equation gives for the Polyakov loop 
2-point function (also refered to as the structure function)
\begin{eqnarray}
\left\langle \widetilde{\psi }(k)\widetilde{\psi }(-k)\right\rangle 
=\left| \widetilde{\psi }(k,0)\right| ^{2}e^{-2\alpha (k^{2}+m^{2})t} \nonumber \\
+\frac{T}{k^{2}+m^{2}}\left( 1-e^{-2\alpha (k^{2}+m^{2})t}\right) ,
\end{eqnarray}
where \( \alpha =4\Gamma T^{2}/g^{2}. \)

When the equation is linearized about \( \psi =0 \) the mass squared is negative
\( m^{2}=-k_{c}^{2}=-g^{2}T^{2}/3 \) causing exponential growth for values of the 
wavenumber less than the critical value
\( k_{c} \). At late times, the equation is linearized about 
the non-trivial minimum at \( \psi =2\pi /3 \). Here the mass is
the Debye screening mass \( m_{D}=gT, \) and the system relaxes to the non-trivial
minimum.

\section{Method}

\subsection{The Bare Potential}

After integrating out all other fields at the scale \( a \) we are left with
an action just in terms of the Polyakov loop.  The minimal Landau-Ginsberg
action which is gauge invariant, $Z(3)$ invariant, and only involves terms up to
the fourth power in the Polyakov loop is
\begin{eqnarray}
S=\int d^{3}x ( |\nabla P_{F}|^{2}+a_{2}(P_{F}^{*}P_{F}) \nonumber \\
+\frac{a_{4}}{2}(P_{F}^{*}P_{F})^{2}+\frac{b_{3}}{3}(P_{F}^{3}+(P_{F}^{*})^{3}) ) ,
\end{eqnarray}
where \( P_{F}=tr_{F}(P) \) is the trace of the Polyakov loop in the fundamental
representation. The Langevin equation without noise is then used as an ansatz
for the evolution of the low momentum modes. In Fourier space this yields an
equation of the form
\begin{eqnarray}
\frac{d}{dt}{\cal F}[P_{F}]
=c_{1}(k^{2}){\cal F}[P_{F}] \nonumber \\ +c_{2}(k^{2}){\cal F}[(P_{F}^{*}P_{F})P_{F}]
+c_{3}(k^{2}){\cal F}[(P_{F}^{*})^{2}] .
\end{eqnarray}

The behavior of the lattice simulation is fit to the above
equation for each \( k \) mode yielding the parameters.
A fit of \( c_{1}(k^{2}) \) versus \( k^{2} \) gives
the critical wavenumber \( a_{2}=-k_{c}^{2} \).

\subsection{The Effective Potential}

The potential obtained above has a minimum which is not the
equilibrium value of the vacuum expectation value of the Polyakov loop(see figure 2). 
This is the bare potential defined at the scale \( a \), while the
effective potential can be defined as the generator at zero momentum of 1PI vertices.
To find the effective potential, the behavior of the 1-point function is fit
to the ansatz
\begin{eqnarray}
V(P_{cl})=a_{2}|P_{cl}|^{2}+\frac{a_{4}}{2}|P_{cl}|^{4} \nonumber \\
+\frac{b_{3}}{3}\left( P_{cl}^{3}+(P_{cl}^{*})^{3}\right) 
\end{eqnarray}
where \( P_{cl}=\left\langle P\right\rangle  \).

Without multiple momentum modes to fit, the absolute normalization
of the coefficients of the effective potential can not be determined. However
the ratio of the curvature at the nontrivial minimum to the curvature at zero
is equal to the ratio of the square of the Debye mass to the square of the critical
wavenumber 
\begin{equation}
\frac{V''(P_{min})}{V''(0)}=\frac{m_{D}^{2}}{k_{c}^{2}}\sim 3,
\end{equation}
which is comparable to the perturbative result for this ratio. The minimum of the
effective potential is located at the final equilibrium value.

\section{Results}

We start from equilibrated lattices of size \( 64^{3}\times 4 \) at \( \beta =5.5 \)
and change the value of \( \beta  \) to a range of values above the deconfinement
value. We use the Cabbibo-Marinari heat bath algorithm
both for equilibration and for the evolution. The large lattice sizes were required
to resolve the long wavelength modes, and to reduce the effects of changing the 
physical volume. 

Quenches to just above the deconfinement temperature show evidence of metastability.  However
more extreme changes in the temperature yield behavior consistent with spinodal decomposition.
\begin{figure}[htb]
\includegraphics[scale=0.30]{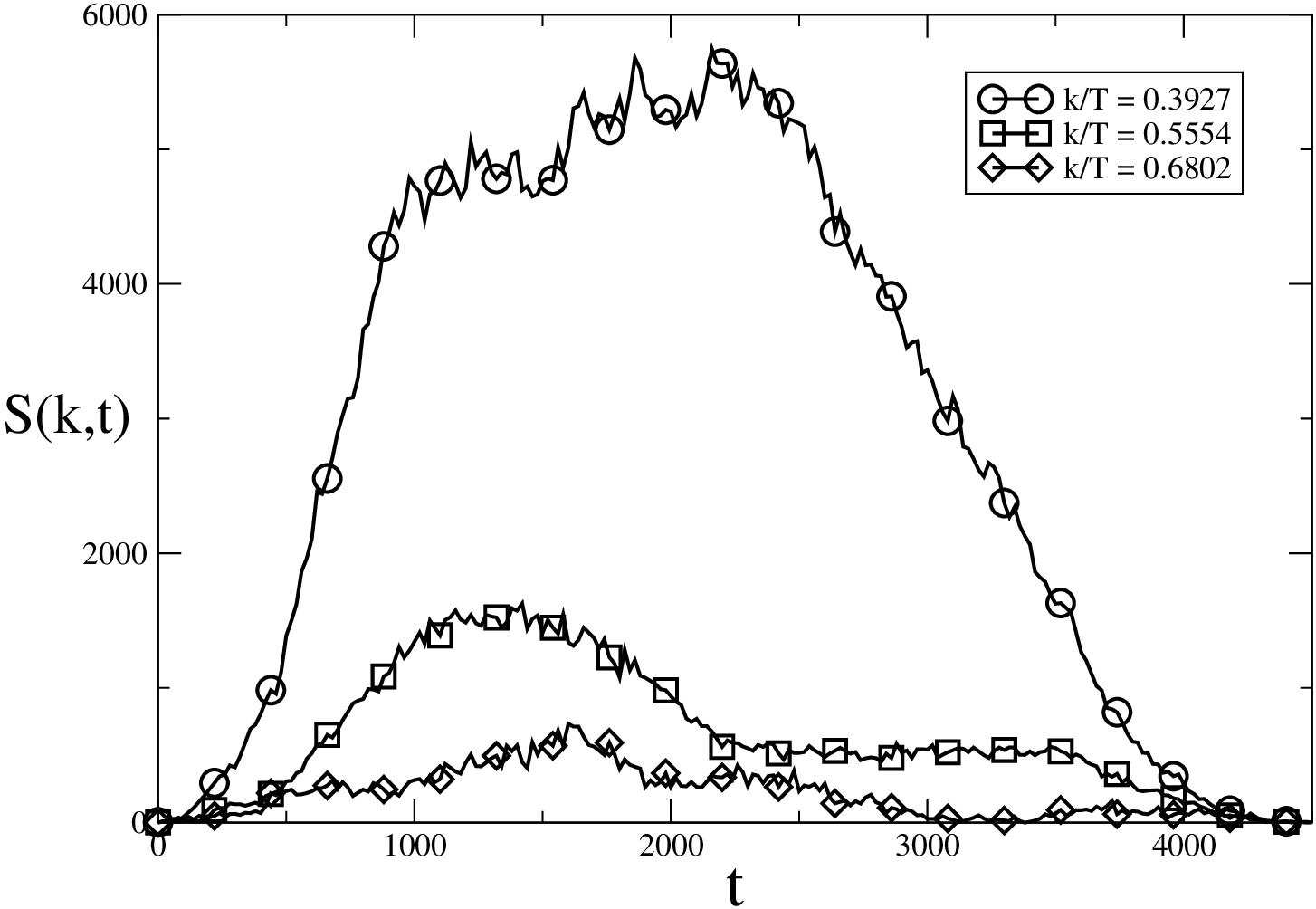}
\vspace{-30pt}
\caption{$S(k,\tau)$ versus Monte Carlo time.}
\label{fig:1}
\vspace{-20pt}
\end{figure}
Typical behavior of the structure function \( S(k,t) \) after a quench to $\beta=6.0$ for 
the three lowest non-trivial wavenumbers is displayed in figure 1 for a single run. The initial 
exponential rise in the structure function is evident. At intermediate times the growth
levels off, with the appearance of ``shelves'' in some of the modes which
is presumably a consequence of mode-mode coupling. At late times the modes relax
to their final equilibrium values.

\begin{figure}[htb]
\vspace{-3pt}
\includegraphics[scale=0.32]{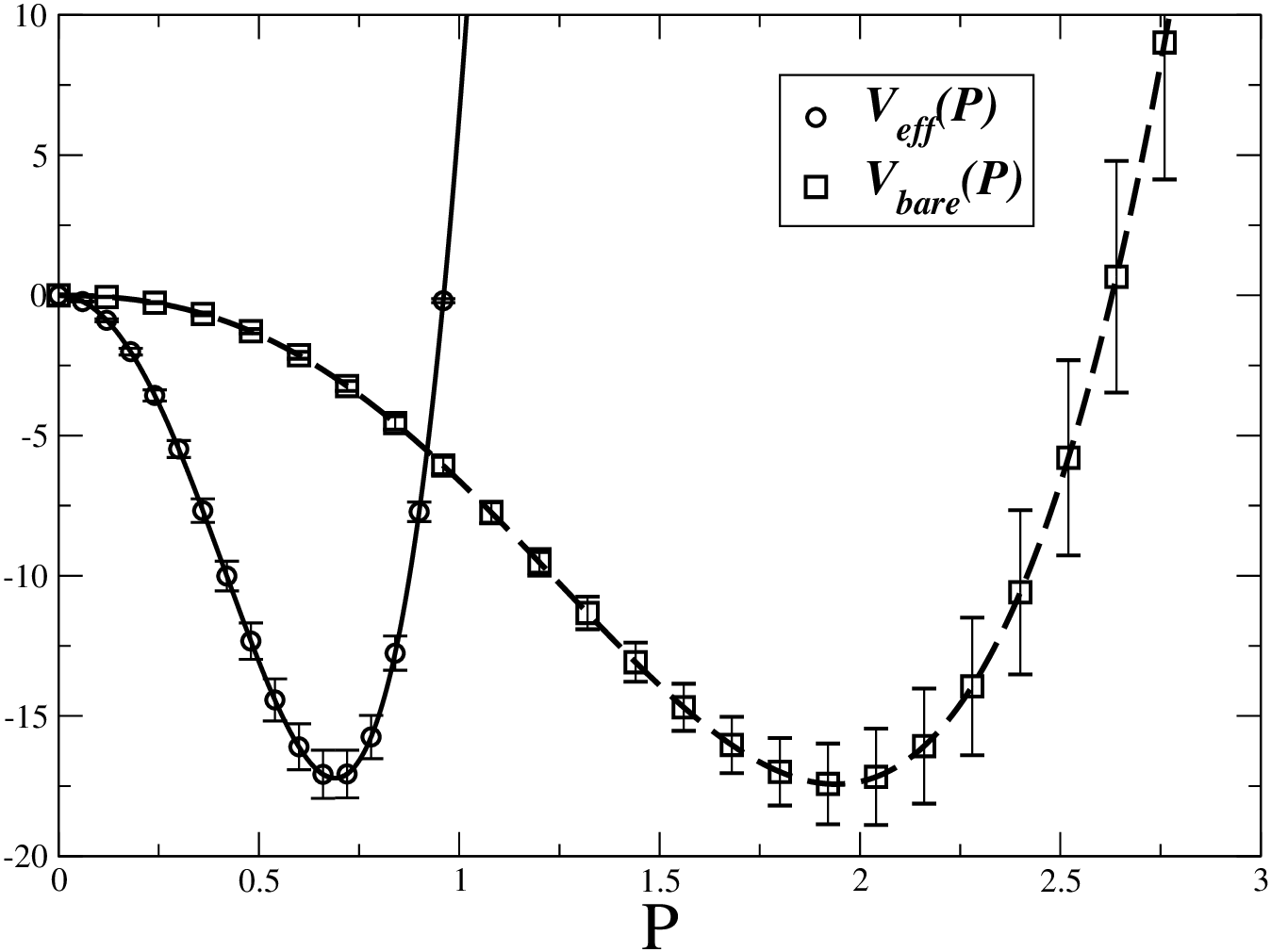}
\vspace{-30pt}
\caption{$V(P)$ vs. $P$}
\label{fig:2}
\vspace{-30pt}
\end{figure}

In figure 2 the bare and the effective potential are plotted. As the
overall normalization of the effective potential cannot be fixed by the present
methods, the normalization was chosen so that both curves could be plotted side
by side. These potentials were obtained by averaging over several runs and
the error bars were estimated using a jackknife analysis. Note the significant
shift in the location of the minimum from the bare to the effective potential. This indicates
that fluctuations are large even at temperatures of the order of $2T_d$.

In figure 3 the value of the critical wavenumber determined from the bare potential
is plotted versus temperature for the four temperatures considered. The temperature
dependence is fit to the ansatz $k_c^2 = aT^2 + bT_d^2,$
which is a form suggested by a model for the pure gauge thermodynamics\cite{us2}. The
fit is used to determine the spinodal temperature, which is defined by 
\( k_{c}(T_{s})=0 \).
The spinodal temperature found is \( T_{s}=1.29(7)T_{d} \) and corresponds
to \( \beta _{s}\approx 5.81 \).

\begin{figure}[htb]
\includegraphics[scale=0.30]{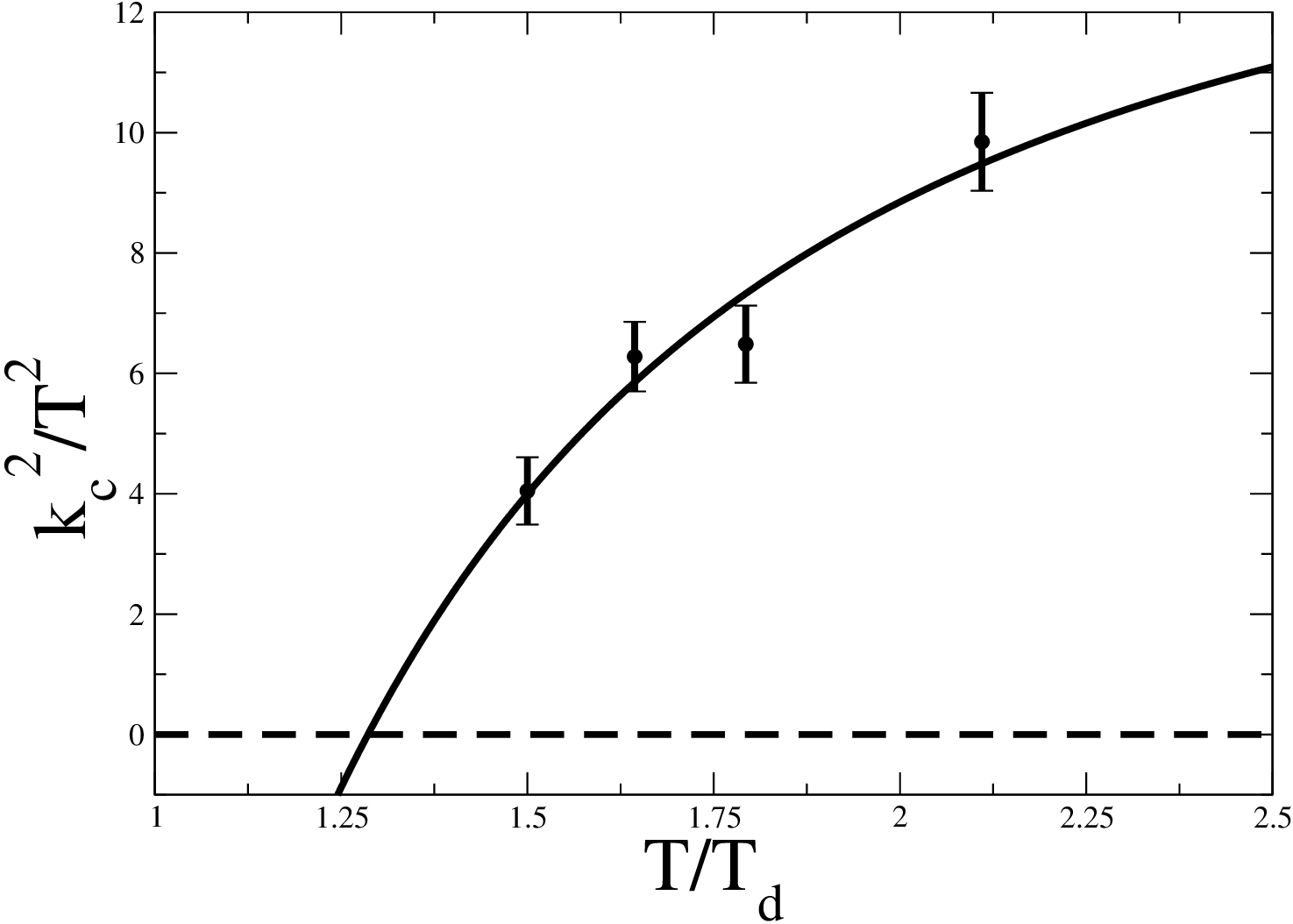}
\vspace{-40pt}
\caption{$k_c(T)$ vs. $T/T_d$}
\label{fig:3}
\vspace{-20pt}
\end{figure}

\section{Conclusion}

Pure $SU(3)$ gauge theory exhibits both metastability and spinodal decomposition
following a quench to temperatures above the deconfinement temperature.  In the spinodal
region the evolution can be modeled using Langevin dynamics, and the bare and effective 
potentials for the Polyakov loop field can be constructed.  These 
potentials determine $k_c$ and $m_D/k_c$.     
We find a nontrivial temperature dependence for the critical wavenumber.
As the temperature is lowered the critical wave number decreases. The extrapolation 
to zero of $k_c(T)$ gives a spinodal temperature slightly above the deconfinement 
temperature.

\end{document}